\documentclass[aps,prd,twocolumn,superscriptaddress,preprintnumbers]{revtex4-2}
\usepackage[colorlinks=true, linkcolor=blue, citecolor=blue, urlcolor=blue]{hyperref}
\usepackage{bm,latexsym,amssymb,amsmath,mathrsfs}
\usepackage{graphicx}
\usepackage{color}
\usepackage{times,setspace}

\begin{document}

\title{Chiral fermion in the Hamiltonian lattice gauge theory}
\author{Tomoya Hayata}
\affiliation{Departments of Physics, Keio University, 4-1-1 Hiyoshi, Kanagawa 223-8521, Japan}
\affiliation{RIKEN iTHEMS, RIKEN, Wako 351-0198, Japan}
\author{Katsumasa Nakayama}
\affiliation{RIKEN Center for Computational Science, Kobe 650-0047, Japan}
\author{Arata Yamamoto}
\affiliation{Department of Physics, The University of Tokyo, Tokyo 113-0033, Japan}
\preprint{RIKEN-iTHEMS-Report-23}

\begin{abstract}
We discuss the chiral fermion in the Hamiltonian formalism of lattice gauge theory.
Although the naive chiral charge operator does not commute with the Hamiltonian,
the commutable one can be defined for the overlap fermion.
The eigenvalues of the energy and the chiral charge can be defined simultaneously.
We study how the eigenvalue spectrum reflects chiral properties of systems, such as a chiral chemical potential and the axial anomaly.
We also show that the Wilson fermion is a chiral fermion in one dimension.
\end{abstract}

\maketitle

\section{Introduction}
The Hamiltonian lattice gauge theory, i.e., the canonical formalism of lattice gauge theory, has a long history.
It started at the dawn of lattice gauge theory \cite{Kogut:1974ag,RevModPhys.51.659}.
Afterwards, however, the path integral formalism~\cite{PhysRevD.10.2445} became mainstream due to its theoretical simplicity and computational advantage.
The Hamiltonian lattice gauge theory was left behind and not fully developed.
Nowadays, there is a revival of the Hamiltonian lattice gauge theory from the viewpoint of quantum computation and tensor network theory to simulate lattice gauge theory without suffering from the notorious sign problem~\cite{Banuls:2019rao,Banuls:2019bmf}.
It would be time to restart the development of the Hamiltonian lattice gauge theory.

In this paper, we focus on the chiral fermion in the Hamiltonian lattice gauge theory.
The Nielsen-Ninomiya no-go theorem prohibits the existence of naive chiral symmetry on the lattice \cite{Nielsen:1980rz,Nielsen:1981xu}.
The solutions to this problem, the so-called chiral fermion, are well established in the path integral formalism.
A famous one is the overlap fermion \cite{Neuberger:1998wv}.
It satisfies the Ginsparg-Wilson relation~\cite{PhysRevD.25.2649} without approximation and holds the chiral symmetry on the lattice.
The overlap fermion is used for the lattice study of chiral phenomena, such as the axial anomaly~\cite{JLQCD:2008jiv,Aoki:2012yj,Cossu:2013uua} and the index theorem \cite{Hasenfratz:1998ri,Luscher:1998pqa,Fujikawa:1998if,Suzuki:1998yz,Adams:1998eg}.
Since the overlap fermion respects the Ginsparg-Wilson relation exactly, we can investigate the chiral properties without any ambiguity.
The discussion at the same level must be possible in the Hamiltonian formalism.
So far, however, we have only limited knowledge of the chiral fermion in the Hamiltonian lattice gauge theory \cite{Horvath:1998gq,Creutz:2001wp,Matsui:2005uh,Mace:2016shq}.
We need to investigate its property in more detail.

This paper is organized as follows.
We start with the definition of the chiral charge operator in the Hamiltonian lattice gauge theory in Sec.~\ref{sec2}.
We introduce the overlap fermion in Sec.~\ref{sec3} and analyze its eigenvalue spectrum on the three-dimensional lattice in Sec.~\ref{sec4}.
Finally, we comment on the one-dimensional case in Sec.~\ref{sec5}.

\section{Hamiltonian lattice gauge theory}\label{sec2}

We consider a $d$-dimensional lattice ($d=1$ or $3$) with periodic boundary conditions.
We use the lattice unit and eliminate the lattice spacing in the following equations.
On the lattice, the fermion creation and annihilation operators satisfy the canonical anti-commutation relation
\begin{equation}
\label{eqacr}
 \left\{ \hat{\psi}_\alpha({\bm x}) , \hat{\psi}^\dagger_\beta({\bm x}) \right\} =  \delta_{\alpha\beta}.
\end{equation}
The indices $\alpha$ and $\beta$ run over flavor, color, and spinor spaces.
The Hamiltonian of a massless Dirac fermion is written in a shorthand notation
\begin{equation}
\label{eqHf}
 \hat{H} = \hat{\psi}^\dagger \gamma^0 \hat{D} \hat{\psi},
\end{equation}
where the coordinate indices, as well as flavor, color, and spinor indices, are implicitly contracted.
Note that $\hat{D}$ is the Dirac operator on the $d$-dimensional space, not on the $(1+d)$-dimensional spacetime.
We put a hat symbol on $\hat{D}$ to emphasize that it contains the quantum link variable operator $\hat{U}_k({\bm x})$ $(k=1,\cdots,d)$.
We use $D$ without a hat symbol when it is a classical matrix.

The Nielsen-Ninomiya no-go theorem says that chiral symmetry is violated on the lattice under several conditions: absence of doublers, translational invariance, Hermiticity, and locality \cite{Nielsen:1980rz,Nielsen:1981xu}.
In the language of the Hamiltonian lattice gauge theory, the lattice fermion cannot satisfy the commutation relation,
\begin{equation}
 [\hat{H},\hat{Q}_{\rm naive}] \neq 0,
\end{equation}
with the naive chiral charge operator
\begin{equation}
 \hat{Q}_{\rm naive} = \hat{\psi}^\dagger \gamma^5 \hat{\psi} .
\end{equation}
As suggested in a pioneering work \cite{Horvath:1998gq}, the definition of the chiral charge operator should be modified by adding a higher-order term of lattice spacing as
\begin{equation}
 \hat{Q} = \hat{\psi}^\dagger \gamma^5 \left(1-\frac{R}{2} \hat{D}\right) \hat{\psi}
\end{equation}
with a real number $R$.
The modified chiral charge operator is required to satisfy the commutation relation
\begin{equation}
 [\hat{H},\hat{Q}] =0
 \label{eqcr1}
 \end{equation}
or equivalently
\begin{equation}
\label{eqcr2}
 \left[ \gamma^0 \hat{D},\gamma^5 \left(1-\frac{R}{2} \hat{D}\right) \right] = 0 .
\end{equation}
This commutation relation defines the notion of the ``chiral fermion'' in the Hamiltonian lattice gauge theory.

\section{Overlap fermion}\label{sec3}

The Hamiltonian formulation of the overlap fermion is constructed by the Dirac operator
\begin{equation}
\label{eqDov}
\hat{D} = \frac{1}{R} \left\{ 1 + \frac{\hat{D}_{\rm W}(m)}{\sqrt{\hat{D}^\dagger_{\rm W}(m)\hat{D}_{\rm W}(m)}} \right\}
\end{equation}
\cite{Creutz:2001wp}.
The kernel is the Wilson-Dirac operator with a negative mass
\begin{equation}
 \hat{D}_{\rm W}(m) = \frac{1}{2} \sum_{k=1}^d \left\{ \gamma^k (\hat{\nabla}_k+\hat{\nabla}_k^\dagger) - \hat{\nabla}_k^\dagger \hat{\nabla}_k \right\} - m
\end{equation}
with the lattice covariant derivative
\begin{eqnarray}
    \hat{\nabla}_k \hat{\psi}({\bm x}) &=& \hat{U}_k({\bm x}) \hat{\psi}({\bm x}+{\bm e}_k) - \hat{\psi}({\bm x}) \\
    \hat{\nabla}_k^\dagger \hat{\psi}({\bm x}) &=& \hat{\psi}({\bm x}) - \hat{U}^\dagger_k({\bm x}-{\bm e}_k) \hat{\psi}({\bm x}-{\bm e}_k)
\end{eqnarray}
where ${\bm e}_k$ is the unit lattice vector in the $x_k$ direction.
The explicit values of $R$ and $m$ are not essential here and a typical choice is $R=m=1$.
The overlap Dirac operator \eqref{eqDov} satisfies the four relations
\begin{align}
 \gamma^5 \hat{D} + \hat{D} \gamma^5 &= R \hat{D} \gamma^5 \hat{D} \label{eqr1} \\
 \gamma^0 \hat{D} + \hat{D} \gamma^0 &= R \hat{D} \gamma^0 \hat{D} \label{eqr2} \\
 \gamma^5 \hat{D} &= \hat{D}^\dagger \gamma^5 \label{eqr3} \\
 \gamma^0 \hat{D} &= \hat{D}^\dagger \gamma^0 \label{eqr4} .
\end{align}
The first one \eqref{eqr1} is the $d$-dimensional version of the Ginsparg-Wilson relation, the second one \eqref{eqr2} is the additional relation in the Hamiltonian formalism, the third one \eqref{eqr3} is called $\gamma^5$-Hermiticity, and the fourth one \eqref{eqr4} is the Hermiticity of the Hamiltonian.
These equations ensure that the overlap fermion satisfies the commutation relation $[\hat{H},\hat{Q}] = 0$.

It is known that the locality of the overlap fermion is ensured only for the admissible (i.e., smooth) gauge configuration in the path integral \cite{Hernandez:1998et}.
In the Hamiltonian formalism, the Dirac operator $\hat{D}$ is a quantum operator acting on the Hilbert space of the gauge field.
When the basis is taken to diagonalize the link variable operators, the matrix is block diagonal and the number of blocks is the number of all possible gauge configurations in the Hilbert space.
The Hilbert space contains not only smooth gauge configurations but also peaky gauge configurations.
Thus the locality is not ensured in the full Hilbert space.
How harmful the non-locality is depends on the problem to solve; for example, it will be harmless for the vacuum property, where low-energy states are dominant, but harmful for real-time dynamics, where all the states are equally relevant.

\section{Eigenvalue spectrum}\label{sec4}

Because of $[\hat{H},\hat{Q}] = 0$, the Hamiltonian and the chiral charge are simultaneously diagonalizable.
Diagonalizing these operators in the full Hilbert space is computationally expensive.
In this paper, we analyze two cases: the free fermion and the fermion coupled with classical background gauge fields.
In both cases, the fermion-fermion interaction mediated by quantum gauge fields is absent.
All we want to know is the eigenvalue spectrum of the one-particle states $|\psi_n\rangle$, s.t.
\begin{equation}
    N = \langle \psi_n| \hat{\psi}^\dagger \hat{\psi} |\psi_n\rangle = 1,
\end{equation}
because an arbitrary multi-particle state is given by a superposition of the one-particle states.
The one-particle Hilbert space $\{|\psi_n\rangle\}$ is much smaller than the full Hilbert space, which contains all the particle number sectors $N=0,1,\cdots,(d+1)L^d$.
The eigenvalues
\begin{align}
\hat{H} |\psi_n\rangle &= \varepsilon_n |\psi_n\rangle \\
\hat{Q} |\psi_n\rangle &= q_n |\psi_n\rangle
\end{align}
are easily obtained by diagonalizing the classical matrices $\gamma^0 D$ and $\gamma^5 \left(1-\frac{R}{2} D\right)$, respectively.
From Eqs.~\eqref{eqr1} to \eqref{eqr4}, we can derive the operator equality
\begin{equation}
    \left( \frac{R}{2} \gamma^0\hat{D}\right)^2 + \left\{ \gamma^5 \left(1-\frac{R}{2} \hat{D}\right) \right\}^2 =1
\end{equation}
and show that the eigenvalues align on a unit circle
\begin{equation}
    \frac{R^2}{4} \varepsilon_n^2 + q_n^2 = 1 .
    \label{eqcircle}
\end{equation}
for all $n$ \cite{Creutz:2001wp}.
This means that the energy and the chirality are correlated on the lattice, while they are independent in the continuous Dirac fermion.

\begin{figure}[t]
\begin{center}
 \includegraphics[width=.45\textwidth]{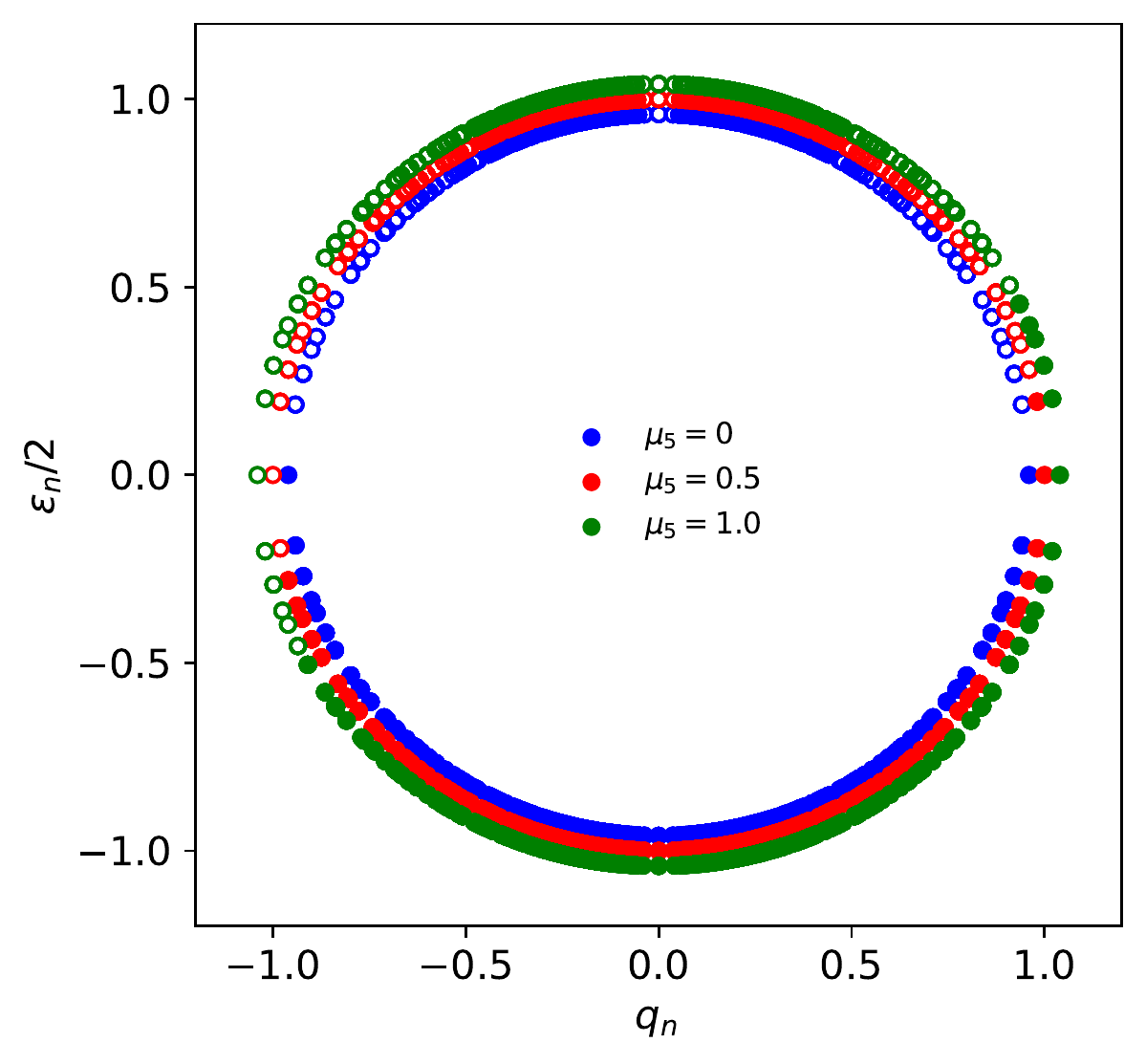}
\caption{
\label{fig1} Eigenvalue spectrum of the free overlap fermion with various chiral chemical potential $\mu_5$.
We set $R=m=1$ and used the $16^3$ lattice with periodic boundary conditions.
Occupied (unoccupied) states are shown by solid (open) circles.
For visibility, the radii of circles are rescaled by hand.
}
\end{center}
\end{figure}

In Fig.~\ref{fig1}, we show the eigenvalue spectrum of the free overlap fermion on the three-dimensional lattice with spatial volume $L^3=16^3$.
We obtained the eigenvalues by numerical diagonalization.
The chirality depends on the energy.
Four zero-energy eigenstates are exactly chiral, $q_n=\pm 1$, and other eigenstates are not.
In the vacuum, all the negative-energy eigenstates are occupied, and their chirality cancels out because of the inversion symmetry $q_n \leftrightarrow - q_n$.
To understand how many zero-energy eigenstates are occupied in the vacuum, we first introduce nonzero mass to Eq.~\eqref{eqDov} and then take the massless limit.
The nonzero mass mixes two left-handed and two right-handed modes, and splits into four linear combinations with zero chirality: two have positive energy and the other two have negative energy.
The latter two are occupied in the vacuum.
Therefore, the total chirality of the vacuum is zero as expected.
The chirality can be externally imbalanced by a chiral chemical potential $\mu_5$~\cite{Fukushima:2008xe}.
Since $\hat{Q}$ is defined as a conserved charge, the spectrum is obtained by shifting as $\hat{H} - \mu_5 \hat{Q}$ (not by changing the kernel \cite{Yamamoto:2011ks}).
The chiral chemical potential tilts the Fermi surface as drawn in Fig.~\ref{fig1}, and makes the total chirality nonzero.
This mimics the dynamical generation of the chirality from the axial anomaly~\cite{NIELSEN1983389,PhysRevB.88.104412}.

\begin{figure}[t]
\begin{center}
 \includegraphics[width=.45\textwidth]{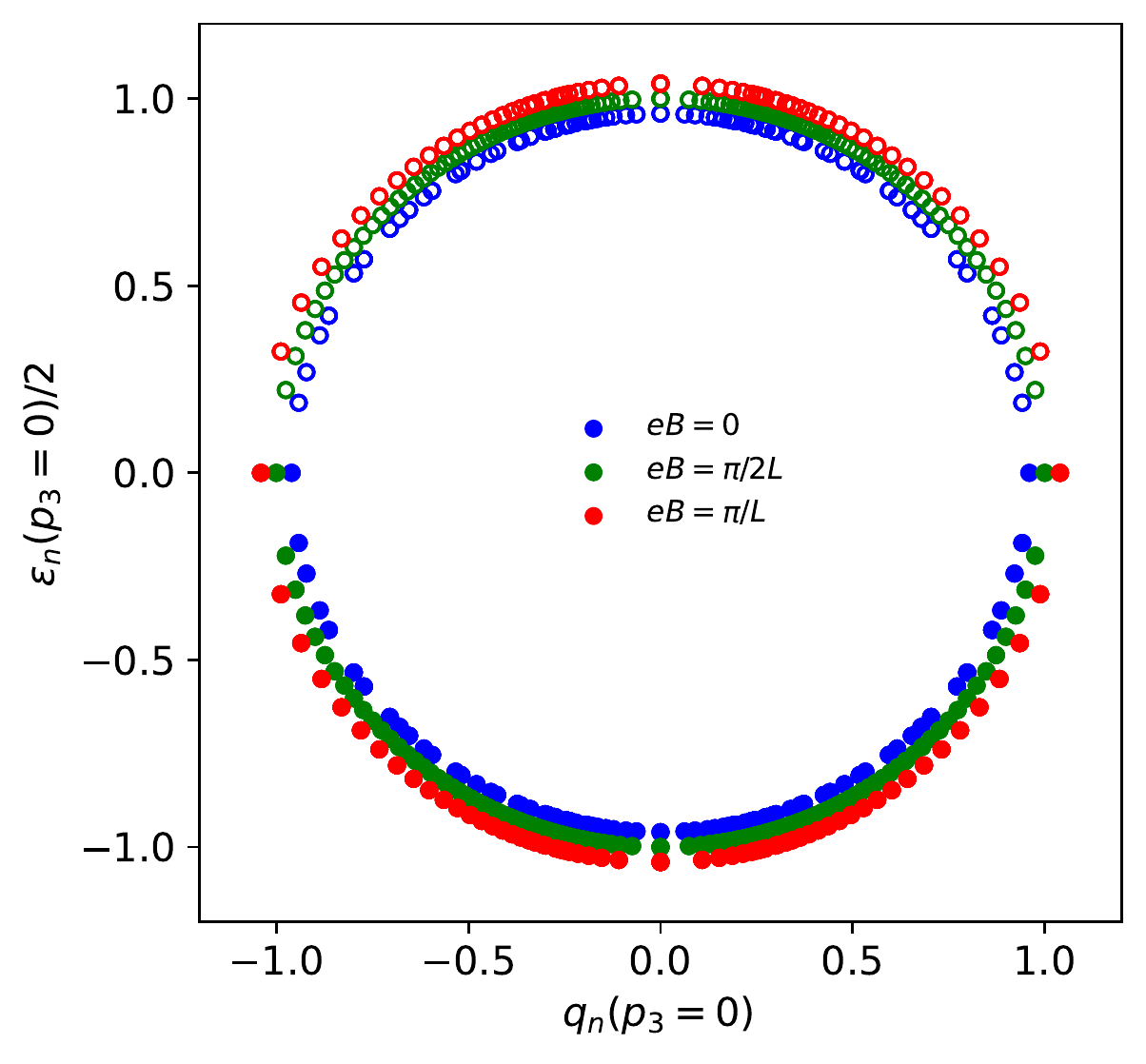}
\caption{
\label{fig2} Eigenvalue spectrum of the overlap fermion with various external magnetic field $B$.
We set $R=m=1$ and used the $16^3$ lattice with periodic boundary conditions.
Occupied (unoccupied) states in the vacuum are shown by solid (open) circles.
To see the gap of the Landau levels clearly, only the $p_3=0$ modes are shown.
For visibility, the radii of circles are rescaled by hand.}
\end{center}
\end{figure}
\begin{figure}[t]
\begin{center}
 \includegraphics[width=.45\textwidth]{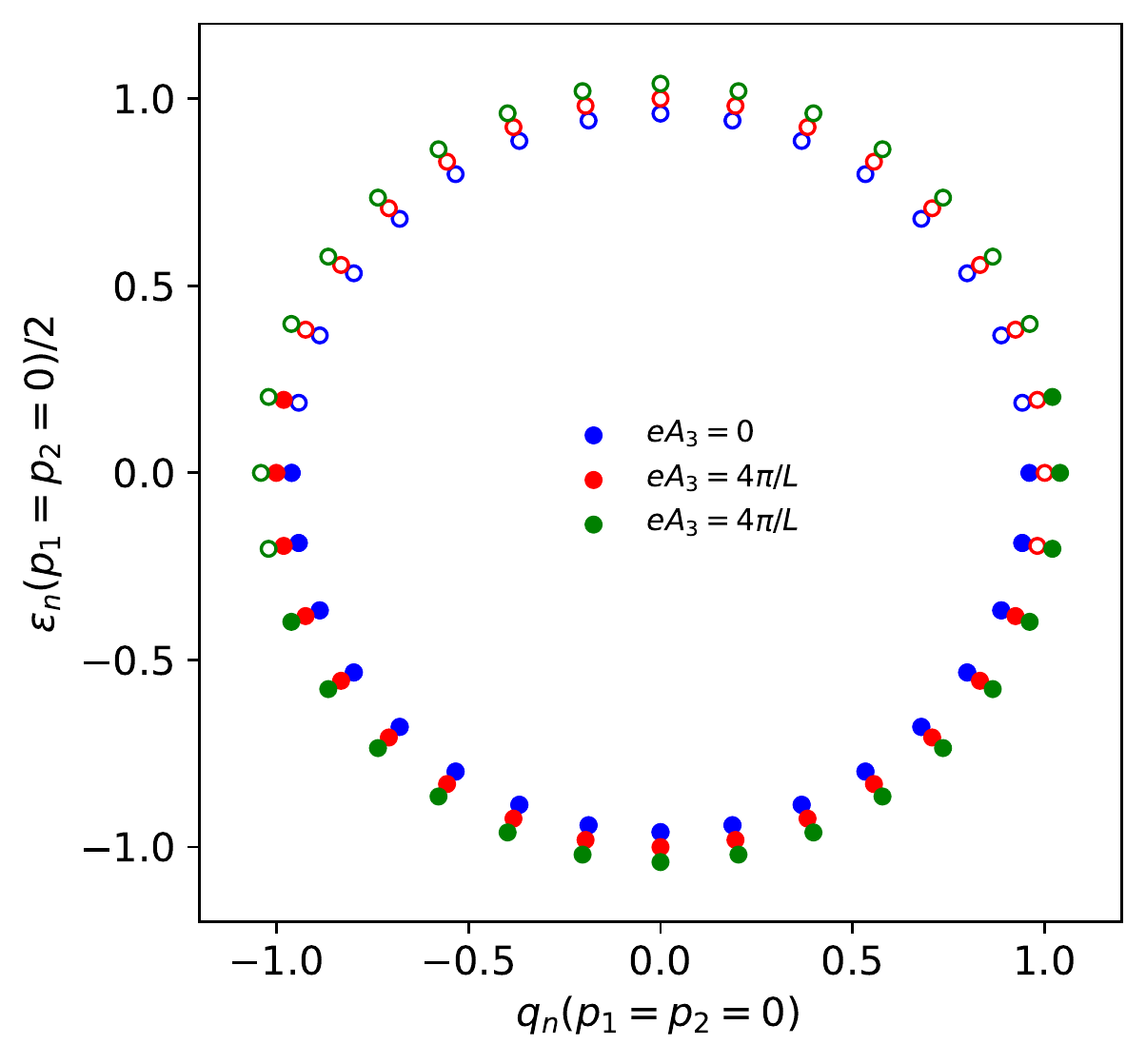}
\caption{
\label{fig3} Spectral flow by external electric field $A_3=Et$.
We set $R=m=1$ and used the $16^3$ lattice with periodic boundary conditions.
Occupied (unoccupied) states under the adiabatic evolution are shown by solid (open) circles.
To see the spectral flow clearly, the two-fold degenerate states after the adiabatic evolution are lifted by hand, which are actually on the same circle with radius one. 
Only the $p_1=p_2=0$ modes are shown for visibility.
Net chirality is not induced solely by the external electric field.
}
\end{center}
\end{figure}
\begin{figure}[t]
\begin{center}
 \includegraphics[width=.45\textwidth]{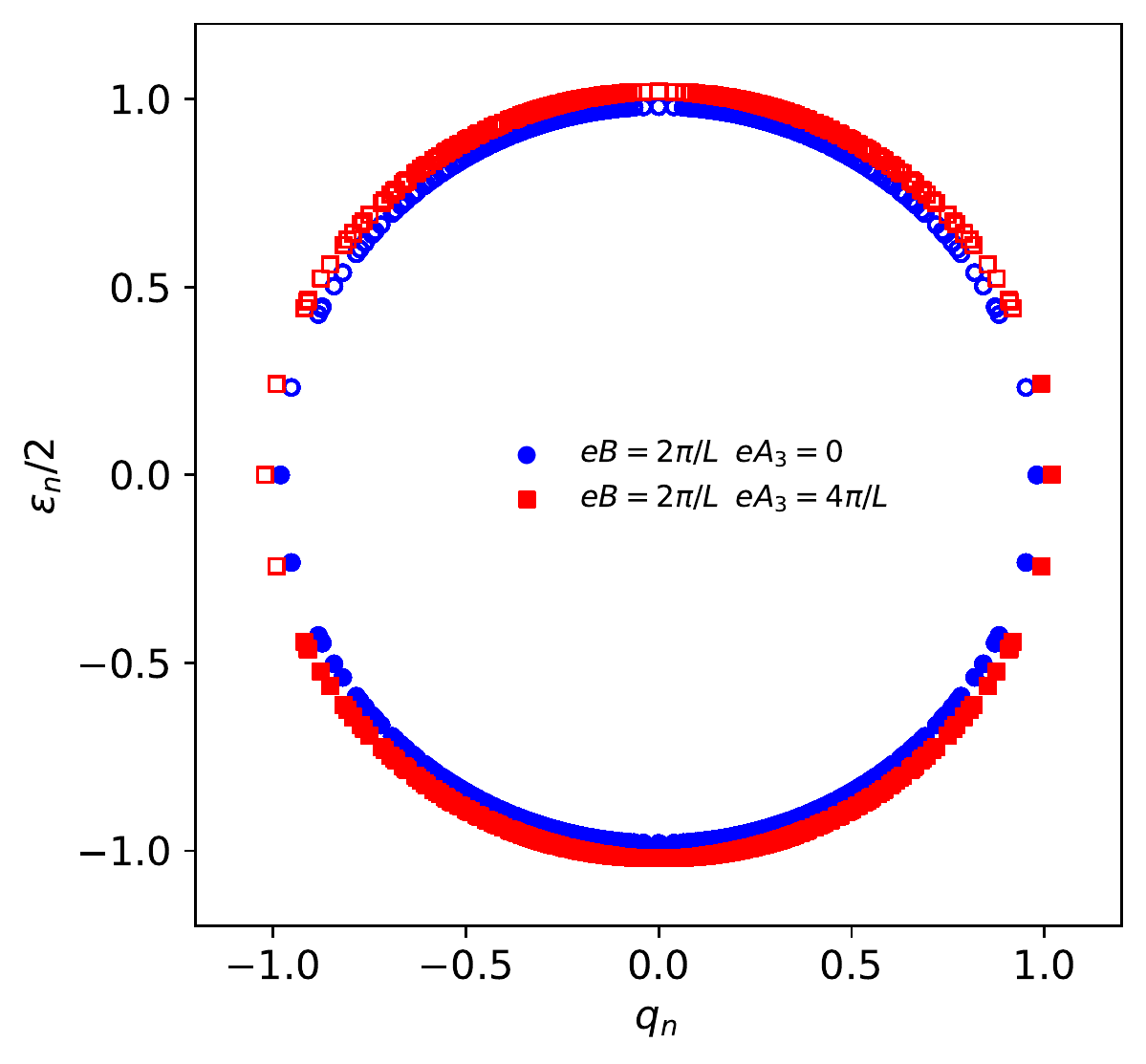}
\caption{
\label{fig4} Spectral flow by external electric and magnetic fields, $eA_3=eEt=4\pi/L$ and $eB=2\pi/L$.
We set $R=m=1$ and used the $16^3$ lattice with periodic boundary conditions.
Occupied (unoccupied) states at the initial state are shown by solid (open) circles, and those after the adiabatic evolution under external electric and magnetic fields are shown by solid (open) squares.
To see the spectral flow clearly, the initial and final states are lifted by hand, which are actually on the same circle with radius one.
Net chirality is dynamically induced by $\bm E\cdot \bm B$.
}
\end{center}
\end{figure}

We next investigate how the spectrum is changed by the classical background of the electromagnetic field.
The classical electromagnetic field is introduced by the c-number link variable
\begin{equation}
    U_k({\bm x})=\exp\{ ieA_k({\bm x}) \} ,
\end{equation}
where $e$ denotes the electric charge.
\begin{description}
\item[Magnetic field]
Let us consider the uniform magnetic field $B$ in the $x_3$ direction.
It can be implemented by setting $A_1({\bm x})=-BL x_2$ at $x_1=L$ and $A_2({\bm x})=Bx_1$  \cite{Al-Hashimi:2008quu}.
The results are shown in Fig.~\ref{fig2}.
The spectral gap increases as $\Delta \varepsilon \sim \sqrt{eB}$ because of the Landau quantization.
The zero-energy chiral modes still exist but they are different from those in the free case.
They are the lowest Landau level; the spin is polarized and the degeneracy is $eBL^2/(2\pi)$.
The magnetic field conserves parity, so the inversion symmetry $q_n \leftrightarrow - q_n$ remains.
\item[Electric field]
Let us consider the uniform electric field $E$ in the $x_3$ direction.
It can be implemented by the time-dependent gauge field $A_3({\bm x})=Et$.
The electric field can be rewritten as the momentum shift $p_3 = 2\pi l/L \to 2\pi l/L + eEt$ ($l=1,\cdots,L$).
When $eEt=4\pi/L$, the shift just moves one momentum to the next-to-next momentum.
Since momentum space is periodic on the lattice, the global spectrum does not change but which states are occupied changes in time.
The occupied states before and after the adiabatic evolution are shown in Fig.~\ref{fig3}. Two degenerate branches (i.e., two spins) differently evolve in $t>0$.
Left-handed modes are favored in one branch and right-handed modes are favored in the other branch.
They are symmetric, so the total chirality is conserved.
\item[Both electric and magnetic fields]
When electric and magnetic fields exist parallelly, a nontrivial thing happens, as shown in Fig.~\ref{fig4}.
Since the spin of the lowest Landau level is polarized along the magnetic field, the chirality locks the direction of motion: for $e>0$, right-handed modes move to the $+x_3$ direction and left-handed modes move to the $-x_3$ direction.
In short, one branch in Fig.~\ref{fig3} is missing.
This breaks the inversion symmetry $q_n \leftrightarrow - q_n$ and generates nonzero net chirality.
This is nothing but dynamical chirality generation by the axial anomaly.
The resultant spectrum is indeed similar to the Fermi surface tilted by the chiral chemical potential in Fig.~\ref{fig1}.
(Strictly speaking, this is not the ``genuine'' anomaly originating from the ultraviolet divergence.
The lattice gauge theory is regularized and the ultraviolet divergence appears only in taking the continuum limit.
A great benefit of the overlap fermion is that the ``would-be'' anomaly can be seen on the regularized spectrum.)
\end{description}

The above investigation can be extended to the case of the quantum gauge field.
When the gauge field is a quantum operator, there are two important differences.
One is that the total Hamiltonian is given by the sum of the gauge field Hamiltonian and the fermion Hamiltonian.
The chiral charge $\hat{Q}$ commutes with the fermion Hamiltonian but does not commute with the gauge field Hamiltonian due to the canonical commutation relation of the gauge field; e.g., $[\hat{E}_k({\bm x}), \hat{U}_k({\bm x})]=e\hat{U}_k({\bm x})$ in the U(1) lattice gauge theory.
The conservation of the chiral charge is anomalously violated.
The unit circle relation \eqref{eqcircle} hold for the eigenstates of the fermion Hamiltonian, but not for the eigenstates of the total Hamiltonian.
The other is computational cost.
The Dirac operator $\hat{D}$ is a quantum operator acting on the Hilbert space of the gauge field.
Its matrix dimension is so huge (exponentially large) that the computational cost exceeds the limit of our current resource.
We have to rely on drastic approximation or wait for technological evolution.

\section{One dimension}\label{sec5}
Finally, we comment on the one-dimensional case, which will be important for benchmark tests of quantum computation and tensor network calculation~\cite{Banuls:2019bmf,Banuls:2019rao}.
The above formulation of the chiral fermion is possible, but there is a more simplified formulation.
Surprisingly enough, the Wilson fermion is equivalent to the overlap fermion~\cite{Horvath:1998gq}.
In one dimension, the Wilson-Dirac operator is
\begin{equation}
 \hat{D}_{\rm W}(m) = \frac{1}{2} \left\{ \gamma^1 (\hat{\nabla}_1+\hat{\nabla}_1^\dagger) - \hat{\nabla}_1^\dagger \hat{\nabla}_1 \right\} - m
\end{equation}
and the gamma matrices $\{\gamma^0,\gamma^1,\gamma^5\}$ are the Pauli matrices.
We can straightforwardly show $\hat{D}^\dagger_{\rm W}(m=1)\hat{D}_{\rm W}(m=1)=1$ and then
\begin{equation}
\begin{split}
 \hat{D} &= 1 + \frac{\hat{D}_{\rm W}(m=1)}{\sqrt{\hat{D}^\dagger_{\rm W}(m=1)\hat{D}_{\rm W}(m=1)}} \\
 &= \hat{D}_{\rm W}(m=0) .
\end{split}
\end{equation}
The massless Wilson-Dirac operator is equal to the overlap Dirac operator with $R=1$.
Thus it satisfies the commutation relation
\begin{equation}
 \left[ \gamma^0 \hat{D}_{\rm W}(m=0),\gamma^5 \left(1-\frac{1}{2} \hat{D}_{\rm W}(m=0)\right) \right] = 0
\end{equation}
or $[\hat{H},\hat{Q}] = 0$.
This is a specialty of the Hamiltonian lattice gauge theory.
In the path-integral formalism, the lowest nontrivial dimension of gauge theory is $1+1$ dimension and the $(1+1)$-dimensional Wilson fermion is not equivalent to the overlap fermion.
In the one-dimensional Hamiltonian lattice gauge theory, the Wilson fermion is a chiral fermion.
This might be counterintuitive because the Wilson fermion is usually said to violate chiral symmetry.
We can study the chiral phenomena of the Wilson fermion just by replacing the naive chiral charge $\hat{Q}_{\rm naive}$ by the modified chiral charge $\hat{Q}$.

\begin{acknowledgments}
The authors thank Yuya Tanizaki for a fruitful discussion.
The authors also thank the summer school on ``A novel numerical approach to quantum field theories" at the Yukawa Institute for Theoretical Physics (YITP-W-22-13) for giving them an opportunity to start this work.
T.~H.~was supported by JSPS KAKENHI Grant No.~~21H01007, and~21H01084.
A.~Y.~was supported by JSPS KAKENHI Grant No.~19K03841.
\end{acknowledgments}

\bibliographystyle{apsrev4-2}
\bibliography{paper}

\end{document}